\documentclass[a4paper,11pt]{article}
\usepackage{jcappub}

\usepackage{mathtools}
\usepackage{pifont}

\newcommand{\gs}{g_\star}
\newcommand{\gss}{g_{\star s}}
\newcommand{\Trh}{T_\text{rh}}
\newcommand{\arh}{a_\text{rh}}
\newcommand{\Tmax}{T_\text{max}}
\newcommand{\amax}{a_\text{max}}
\newcommand{\rGW}{\rho_\text{GW}}
\newcommand{\rR}{\rho_R}
\newcommand{\rp}{\rho_\phi}
\newcommand{\oGW}{\Omega_\text{GW}}
\newcommand{\DNeff}{\Delta N_\text{eff}}
\newcommand{\Gp}{\Gamma_\phi}
\newcommand{\Mrh}{M_\text{rh}}
\newcommand{\mueff}{\mu_\text{eff}}
\newcommand{\yeff}{y_\text{eff}}
\newcommand{\Eom}{E_\omega}

\title{Bremsstrahlung-induced\\Gravitational Waves\\in Monomial Potentials\\during Reheating}
\author[a]{Basabendu Barman,}
\author[b]{Nicolás Bernal,}
\author[c]{\\Yong Xu,}
\author[d]{and Óscar Zapata}

\affiliation[a]{Institute of Theoretical Physics, Faculty of Physics, University of Warsaw\\
	ul. Pasteura 5, 02-093 Warsaw, Poland}
\affiliation[b]{New York University Abu Dhabi\\
	PO Box 129188, Saadiyat Island, Abu Dhabi, United Arab Emirates}
\affiliation[c]{PRISMA+ Cluster of Excellence and Mainz Institute for Theoretical Physics\\
	Johannes Gutenberg University, 55099 Mainz, Germany}
\affiliation[d]{Instituto de Física, Universidad de Antioquia\\
	Calle 70 \# 52-21, Apartado Aéreo 1226, Medellín, Colombia}

\emailAdd{basabendu88barman@gmail.com}
\emailAdd{nicolas.bernal@nyu.edu}
\emailAdd{yonxu@uni-mainz.de}
\emailAdd{oalberto.zapata@udea.edu.co}

\abstract{
	We discuss the production of primordial gravitational waves (GW) from radiative inflaton decay during the period of reheating, assuming perturbative decay of the inflaton either into a pair of bosons or fermions, leading to successful reheating satisfying constraint from Big Bang nucleosynthesis. Assuming that the inflaton $\phi$ oscillates in a general monomial potential $V(\phi)\propto \phi^n$, which results in a time-dependent inflaton decay width, we show that the resulting stochastic GW background can have optimistic detection prospects, especially in detectors that search for a high-frequency GW spectrum, depending on the choice of $n$ that determines the shape of the potential during reheating. We also discuss how this GW energy density may affect the measurement of $\DNeff$ for bosonic and fermionic reheating scenarios.
}

\begin{document}
	\begin{flushright}
	\end{flushright}
	\maketitle
	
	\section{Introduction}
	The primordial gravitational wave (GW) is an essential cosmic relic that can provide direct insight into the unknown physics of the early universe. It can potentially originate from various processes, {\it viz.} quantum fluctuations during inflation, particle production during preheating, or oscillations of cosmic string loops, as discussed, for example, in Refs.~\cite{Maggiore:2018sht, Caprini:2018mtu}. Recently,  Refs.~\cite{Nakayama:2018ptw, Huang:2019lgd, Ghoshal:2022kqp, Barman:2023ymn} have highlighted the possibility of a stochastic GW background generated by the radiative decay of inflatons during the reheating period. This phenomenon results from the irreducible interaction of gravitons, quantum fluctuations emerging over the classical background, with all matter fields. As a consequence, graviton bremsstrahlung becomes an inevitable process.\footnote{We notice that thermal fluctuations in the Standard Model (SM) plasma could trigger a similar process~\cite{Ghiglieri:2015nfa, Ghiglieri:2020mhm, Ghiglieri:2022rfp}.} Although the amplitude of this GW spectrum is suppressed by the Planck scale due to the minimal coupling with gravity, it could still be detected in future high-frequency GW detectors such as resonant cavities~\cite{Berlin:2021txa, Herman:2022fau, Berlin:2023grv}, as illustrated in Ref.~\cite{Barman:2023ymn}. We refer the reader to Ref.~\cite{Aggarwal:2020olq} for a recent review on the detection prospects of high-frequency GWs.
	
	It is generally assumed that the (p)reheating period~\cite{Kofman:1997yn} is preceded by a cosmic inflationary epoch~\cite{Starobinsky:1980te, Guth:1980zm, Linde:1981mu, Albrecht:1982wi} in which the inflaton $\phi$ slowly rolls along a plateau on its way to the minimum of the scalar potential. The cosmic reheating period (coherent oscillations around the minimum) can be approximated by different inflaton potentials $V(\phi)$.\footnote{Notice that the slow-roll inflationary models $\phi^n$ with $n = 2\,,4$ that are assumed to be valid for both inflation and reheating periods are disfavored by Planck constraints~\cite{Planck:2018jri} on the scalar index $n_s$ and the tensor-to-scalar ratio $r$ for cosmological density perturbations.} In this article, we go beyond the previous results in Refs.~\cite{Nakayama:2018ptw, Huang:2019lgd, Barman:2023ymn}, and consider general monomial potentials $\phi^n$ for the inflaton during reheating. Such potentials can naturally arise from, for example, the $\alpha$-attractor $T$- and $E$-model~\cite{Kallosh:2013hoa, Kallosh:2013yoa, Kallosh:2013maa} or the Starobinsky inflationary model~\cite{Starobinsky:1980te, Starobinsky:1981vz, Starobinsky:1983zz, Kofman:1985aw}.
	
	As the inflaton oscillates in a monomial potential $\phi^n$ during reheating, its equation of state (EoS) $w = (n-2)/(n+2)$ depends on the shape of the potential, while the decay width of the inflaton develops a time dependence for $n \neq 2$~\cite{Garcia:2020eof, Garcia:2020wiy, Ahmed:2021fvt, Barman:2022tzk, Ahmed:2022tfm, Bernal:2022wck, Chakraborty:2023ocr, Barman:2023ktz}. We consider the inflaton to perturbatively decay into either bosonic or fermionic final states, and in each case, we compute the GW energy density originating from the graviton bremsstrahlung process. We find that the detection prospect of such high-frequency stochastic GW is crucially dependent on the choice of $n$. For the bosonic reheating scenario, the spectrum can be significantly boosted for $n>2$, while such a boost is less prominent in the case of fermionic reheating. We also compute the upper bound on the GW energy density at the time of Big Bang nucleosynthesis (BBN) and the cosmic microwave background (CMB) decoupling, in terms of $\DNeff$, which constrains the number of relativistic degrees of freedom tolerated at those cosmic epochs. However, we find that the current or even future measurements of $\DNeff$ hardly put any bound on the GW energy density in the present scenario.  
	
	The paper is organized as follows. In Section~\ref{sec:framework} we briefly go through the underlying model that gives rise to different decay channels. Details of the reheating with a monomial potential are elaborated in Section~\ref{sec:reheating}. We then compute the primordial GW spectrum from gravitational bremsstrahlung during reheating in Section~\ref{sec:gw}, and discuss its detection prospects in Section~\ref{sec: results}. Finally, we summarize our results and draw conclusions in Section~\ref{sec:concl}. 
	
	\section{The Setup} 
	\label{sec:framework}
	Here, we briefly discuss the underlying interaction Lagrangian that gives rise to the relevant decay widths for the inflaton. The particle physics framework of the present analysis is entirely based on the model in Ref.~\cite{Barman:2023ymn}. For two-body decay processes, we consider trilinear interactions between the inflaton $\phi$ and a pair of complex scalar doublets $\varphi$ (e.g. the Higgs boson doublet) or a pair of vector-like Dirac fermions $\Psi$. The corresponding Lagrangian density reads
	\begin{equation}\label{eq:int1}
	\mathcal{L}_\text{int} \supset -\mu\, \phi\, |\varphi|^2 - y_\psi\, \overline{\Psi}\, \Psi\, \phi\,.    
	\end{equation}
	The interaction strengths are parameterized in terms of the couplings $\mu$ and $y_\psi$, respectively. Note that $\mu$ is a dimensionful coupling. On the other hand, when expanding the metric $g_{\mu \nu}$ around a flat spacetime: $g_{\mu \nu} \simeq \eta_{\mu \nu} + \frac{2}{M_P}\, h_{\mu \nu}$, we end up with gravitational bremsstrahlung processes involving massless graviton final states via the interaction of the form~\cite{Choi:1994ax}
	\begin{equation} \label{eq:int2}
	\sqrt{-g}\, \mathcal{L}_{\rm int}^{(g)} \supset -\frac{2}{M_P}\, h_{\mu \nu}\, T^{\mu \nu},
	\end{equation}
	where $h_{\mu\nu}$ refers to the graviton field that appears as a quantum fluctuation on the flat background, and $T_{\mu\nu}$ represents the energy-momentum tensor that involves matter particles in the theory. Here $M_P$ is the reduced Planck mass. The interactions in Eqs.~\eqref{eq:int1} and~\eqref{eq:int2} give rise to 2- and 3-body decays of the inflaton into pairs of $\varphi$ and $\Psi$, along with the radiative emission of a massless graviton. The 3-body decay processes are of present interest, as they are the source of the stochastic GW background. Full expressions of the differential 3-body decay rates are reported in Appendix~\ref{sec:3body}.
	
	\section{Reheating Dynamics}
	\label{sec:reheating}
	In this section, we closely follow Ref.~\cite{Bernal:2022wck} and revisit the details of reheating through bosonic and fermionic decays, considering the evolution of the energy densities of the inflaton and radiation.  We consider the post-inflationary oscillation of the inflaton $\phi$ at the bottom of a monomial potential $V(\phi)$ of the form
	\begin{equation}\label{eq:inf-pot}
	V(\phi) = \lambda\, \frac{\phi^n}{\Lambda^{n - 4}}\,,
	\end{equation}
	where $\lambda$ is a dimensionless coupling and $\Lambda$ an energy scale. Now, the equation of motion for the oscillating inflaton field reads~\cite{Turner:1983he}
	\begin{equation} \label{eq:eom0}
	\ddot\phi + (3\, H + \Gp)\, \dot\phi + V'(\phi) = 0\,,
	\end{equation} 
	where $H$ denotes the Hubble expansion rate, $\Gp$ the inflaton decay rate, dots $(\dot {\phantom .})$ derivatives with respect to time, and primes ($'$) derivatives with respect to the field.
	Defining the energy density and pressure of $\phi$ as $\rp \equiv \frac12\, \dot\phi^2+ V(\phi)$ and $p_\phi \equiv \frac12\, \dot\phi^2 - V(\phi)$, together with the EoS parameter $w \equiv p_\phi/\rp = (n - 2) / (n + 2)$~\cite{Turner:1983he}, one can write the evolution of the inflaton energy density as
	\begin{equation} \label{eq:drhodt}
	\frac{d\rp}{dt} + \frac{6\, n}{2 + n}\, H\, \rp = - \frac{2\, n}{2 + n}\, \Gp\, \rp\,.
	\end{equation}
	During reheating $a_I \ll a \ll \arh$, where $a$ is the scale factor, the term associated with expansion, that is, $H\, \rp$ typically dominates over the interaction term $\Gp\, \rp$. Then it is possible to solve Eq.~\eqref{eq:drhodt} analytically, leading to
	\begin{equation} \label{eq:rpsol}
	\rp(a) \simeq \rp (\arh) \left(\frac{\arh}{a}\right)^\frac{6\, n}{2 + n}.
	\end{equation}
	Here, $a_I$ and $\arh$ correspond to the scale factor at the end of inflation and at the end of reheating, respectively. Since the Hubble rate during reheating is dominated by the inflaton energy density, it follows that
	\begin{equation} \label{eq:Hubble}
	H(a) \simeq H(\arh) \times
	\begin{dcases}
	\left(\frac{\arh}{a}\right)^\frac{3\, n}{n + 2} &\text{ for } a \leq \arh\,,\\
	\left(\frac{\arh}{a}\right)^2 &\text{ for } \arh \leq a\,.
	\end{dcases}
	\end{equation}
	At the end of the reheating (that is, at $a = \arh$), the energy densities of the inflaton and radiation are equal, $\rR(\arh) = \rp(\arh) = 3\, M_P^2\, H(\arh)^2$.
	Note that to avoid affecting the success of BBN, the reheating temperature $\Trh$ must satisfy $\Trh > T_\text{BBN} \simeq 4$~MeV~\cite{Sarkar:1995dd, Kawasaki:2000en, Hannestad:2004px, DeBernardis:2008zz, deSalas:2015glj,Hasegawa:2019jsa}.
	The evolution of the SM radiation energy density $\rR$, on the other hand, is governed by the Boltzmann equation of the form~\cite{Garcia:2020wiy}
	\begin{equation} \label{eq:rR}
	\frac{d\rR}{dt} + 4\, H\, \rR = + \frac{2\, n}{2 + n}\, \Gp\, \rp\,,
	\end{equation}
	where it is implicitly assumed that during reheating the inflaton energy density is transferred to the radiation energy density. Using Eq.~\eqref{eq:rpsol}, one can solve Eq.~\eqref{eq:rR} and further obtain 
	\begin{equation} \label{eq:rR_int}
	\rR(a) \simeq \frac{2\, \sqrt{3}\, n}{2 + n}\, \frac{M_P}{a^4} \int_{a_I}^a \Gp(a')\, \sqrt{\rp(a')}\, a'^3\, da'\,.
	\end{equation}
	Note that here a general scale factor dependence of $\Gp$ has been assumed, which may arise, for example, from the field-dependent inflaton mass. In the present setup, the effective mass $M(a)$ for the inflaton can be obtained from the second derivative of Eq.~\eqref{eq:inf-pot}, which reads
	\begin{equation}\label{eq:inf-mass1}
	M(a)^2 \equiv \frac{d^2V}{d\phi^2} = n\, (n - 1)\, \lambda\, \frac{\phi^{n - 2}}{\Lambda^{n - 4}}
	\simeq n\, (n-1)\, \lambda^\frac{2}{n}\, \Lambda^\frac{2\, (4 - n)}{n} \rp(a)^{\frac{n-2}{n}}\,,
	\end{equation}
	or equivalently,
	\begin{equation}\label{eq:inf-mass}
	M(a) = \Mrh \left(\frac{\arh}{a}\right)^\frac{3\, (n - 2)}{n + 2},
	\end{equation}
	where
	\begin{equation} \label{eq:Mrh}
	\Mrh \equiv M(\arh) \simeq \sqrt{n\, (n-1)}\, \lambda^\frac{1}{n}\, \Lambda^\frac{4 - n}{n} \left[3\, M_P^2\, H^2(\Trh)\right]^{\frac{n-2}{2 n}}.
	\end{equation}
	It is interesting to note that for $n \neq 2$, $M$ has a field dependence that, in turn, would lead to an inflaton decay rate with a scale factor (or time) dependence.
	
	Before moving further, we would like to note that the potential in Eq.~\eqref{eq:inf-pot} can naturally arise in a number of inflationary scenarios, for example, the $\alpha$-attractor T- or E-models~\cite{Kallosh:2013hoa, Kallosh:2013yoa, Kallosh:2013maa}, or the Starobinsky model~\cite{Starobinsky:1980te, Starobinsky:1981vz, Starobinsky:1983zz, Kofman:1985aw}. Now, given a particular inflationary model, for example, in $\alpha$-attractor T-model~\cite{Kallosh:2013hoa, Kallosh:2013yoa}
	\begin{equation}
	V(\phi ) =\lambda\, M_P^4 \left[\tanh \left(\frac{\phi}{\sqrt{6\, \alpha}\, M_P}\right)\right]^n \simeq \lambda\, M_P^4 \times
	\begin{dcases}
	1 & \; \text{for}\; \phi \gg M_P,\\
	\left(\frac{\phi}{\sqrt{6\,\alpha}\,M_P}\right)^n & \; \text{for}\; \phi\ll M_P\,.
	\end{dcases}
	\end{equation}
	The overall scale of the potential parameterized by the coupling $\lambda$ can be determined from the amplitude of the scalar perturbation power spectrum $A_S \simeq (2.1 \pm 0.1) \times 10^{-9}$~\cite{Planck:2018jri},  
	\begin{equation} \label{eq:lam}
	\lambda \simeq \frac{18\,\pi^2\,\alpha\,A_S}{6^{n/2}\, N_\star^2} \,,
	\end{equation}
	where $N_\star$ is the number of $e$-folds measured from the end of inflation to the time when the pivot scale $k_\star = 0.05$~Mpc$^{-1}$ exits the horizon. Furthermore, $\Mrh$ no longer remains a free parameter, as it is fixed by $\lambda$ and $\Trh$.  However, we keep our discussion as general as possible and do not consider any particular underlying inflationary potential. With this premise, we now move on to the discussion of two reheating scenarios, where the reheating is completed via inflaton decays either into a pair of bosons or into a pair of fermions [cf. Eq.~\eqref{eq:int1}].
	
	\subsection{Fermionic reheating}
	We first consider the scenario where the inflaton decays into a pair of fermions via the Yukawa interaction in Eq.~\eqref{eq:int1}, with a decay rate
	\begin{equation} \label{eq:fer_gamma}
	\Gp(a) = \frac{\yeff^2}{8\pi}\, M(a)\,,
	\end{equation}
	where the effective coupling $\yeff \ne y_\psi$ (for $n \neq 2$) is obtained after averaging over several oscillations~\cite{Shtanov:1994ce, Ichikawa:2008ne, Garcia:2020wiy}. The evolution of the SM energy density (Eq.~\eqref{eq:rR_int}) in this case becomes~\cite{Bernal:2022wck}
	\begin{equation} \label{eq:rR_fer}
	\rR(a) \simeq \frac{3\, n}{7 - n}\, M_P^2\, \Gp(\arh)\, H(\arh) \left(\frac{\arh}{a}\right)^\frac{6 (n - 1)}{2 + n} \left[1 - \left(\frac{a_I}{a}\right)^\frac{2 (7 - n)}{2 + n}\right],
	\end{equation}
	and, therefore, the temperature of the SM bath evolves as
	\begin{equation} \label{eq:Tevol}
	T(a) \simeq \Trh \left(\frac{\arh}{a}\right)^\alpha,
	\end{equation}
	with
	\begin{equation} \label{eq:Tfer}
	\alpha =
	\begin{dcases}
	\frac32\, \frac{n - 1}{n + 2} & \text{ for } n < 7\,,\\
	1 & \text{ for } n > 7\,.
	\end{dcases}
	\end{equation}
	Trading the scale factor with temperature, the Hubble expansion rate during reheating (cf. Eq.~\eqref{eq:Hubble}) can be rewritten as
	\begin{equation} \label{eq:Hevol}
	H(T) \simeq H(\Trh) \left(\frac{T}{\Trh}\right)^{\frac{3\, n}{2 + n}\, \frac{1}{\alpha}}.
	\end{equation}
	Note that, for the case with $n = 2$ where the inflaton oscillates in a quadratic potential with an EoS parameter $w = 0$, the standard dependences with the scale factor $\rR(a) \propto a^{-3/2}$ and $T(a) \propto a^{-3/8}$ are reproduced. 
	
	\subsection{Bosonic reheating}
	Alternatively, if inflatons only decay into a pair of bosons through the trilinear scalar interaction in Eq.~\eqref{eq:int1}, the decay rate is instead
	\begin{equation} \label{eq:bos_gamma}
	\Gp(a) = \frac{\mueff^2}{8\pi\, M(a)}\,,
	\end{equation}
	where again the effective coupling $\mueff \ne \mu$ (if $n\neq2$) can be obtained after averaging over oscillations. Using a procedure similar to the previous fermionic case, one sees that the SM energy density scales as~\cite{Bernal:2022wck}
	\begin{equation} \label{eq:rR_bos}
	\rR(a) \simeq \frac{3\, n}{1 + 2\, n}\, M_P^2\, \Gp(\arh)\, H(\arh) \left(\frac{\arh}{a}\right)^\frac{6}{2 + n} \left[1 - \left(\frac{a_I}{a}\right)^\frac{2\, (1 + 2 n)}{2 + n}\right]\,,
	\end{equation}
	with which the SM temperature and the Hubble rate evolve as Eqs.~\eqref{eq:Tevol} and~\eqref{eq:Hevol}, respectively, with
	\begin{equation} \label{eq:TBos}
	\alpha = \frac32\, \frac{1}{n + 2}
	\end{equation}
	during reheating. As before, we reproduce the results of the oscillation in a quadratic potential for $n = 2$. Note that for $n > 2$ the radiation energy density is diluted faster in the case of fermionic reheating than in bosonic reheating. As an example, for $n = 4$, the radiation energy density decreases as $\rR \propto a^{-3}$ for fermionic and $\rR \propto a^{-1}$ for bosonic reheating, respectively. It is important to briefly discuss the qualitative behavior of the two reheating scenarios. As we have already noticed, the inflaton decay width into fermionic final states $\propto M(a)$, while for bosonic final states the decay width $\propto 1/M(a)$. Since the inflaton mass $M(a)$ is a decreasing function of time, the reheating process becomes more efficient over time for the bosonic final states than for the fermionic final states for $n>2$.
	
	Finally, we would like to mention that for $n \gtrsim 8$ (or equivalently $\omega \gtrsim 0.65$), purely gravitational reheating becomes important~\cite{Haque:2022kez, Clery:2022wib, Co:2022bgh, Haque:2023yra} and may dominate over perturbative reheating for certain choices of the inflaton-matter couplings, as shown in Refs.~\cite{Haque:2022kez, Haque:2023yra}. Typically, for $\omega\gtrsim 0.65$, gravitational reheating alone can be sufficient to reheat the Universe, without requiring any contribution from perturbative reheating.\footnote{Such scenarios, however, are severely constrained from inflationary GW overproduction~\cite{Opferkuch:2019zbd, Figueroa:2018twl, Barman:2022qgt}.} Since we are interested in the perturbative reheating scenario, in the subsequent analysis we shall therefore consider $n < 8$.
	
	\section{Gravitational Waves from  Graviton Bremsstrahlung}
	\label{sec:gw}
	After production via the bremsstrahlung process, gravitons propagate and constitute a sto\-chas\-tic GW background. In this section, we compute the contribution to the effective number of neutrinos $N_\text{eff}$ and the GW spectrum, considering bosonic and fermionic reheating scenarios. To do that, we first write down the Boltzmann equations for the inflaton, SM radiation, and GW energy densities as
	\begin{align}
	&\frac{d\rp}{dt} + \frac{6\, n}{2 + n}\, H\, \rp = - \frac{2\, n}{2 + n}\, \left(\Gamma^{(0)} + \Gamma^{(1)}\right) \rp\,, \label{eq:beq1}\\
	&\frac{d\rR}{dt} + 4\, H\, \rR = + \frac{2\, n}{2 + n}\, \Gamma^{(0)}\, \rp + \frac{2\, n}{2 + n}\, \int \frac{d\Gamma^{(1)}}{d\Eom}\, \frac{M - \Eom}{M}\, \rp\, d\Eom\,, \label{eq:beq2}\\
	&\frac{d\rGW}{dt} + 4\, H\, \rGW = + \frac{2\, n}{2 + n}\, \int \frac{d\Gamma^{(1)}}{d\Eom}\, \frac{\Eom}{M}\, \rp\, d\Eom\,, \label{eq:beq3}
	\end{align}
	where $\Eom$ corresponds to the graviton energy at the moment of production, $\Gamma^{(0)}$ to the 2-body inflaton decay width in Eqs.~\eqref{eq:fer_gamma} and~\eqref{eq:bos_gamma}, and $\Gamma^{(1)}$ to the 3-body decay width in Eqs.~\eqref{eq:dGdE_bos} and~\eqref{eq:dGdE_fer}.%
	\footnote{We note that the 2-to-2 process inflaton + $\psi$ $\to$ graviton + $\psi$ (with $\psi$ being a decay product of inflaton) has the same topology as the 3-body decay. However, the contribution of the 2-to-2 process is subdominant compared to the 1-to-3 decay in our case because the inflaton mass is always larger than the bath temperature~\cite{Klose:2022knn}.
	} 
	We note that, for $n = 2$, these equations reproduce those presented in Ref.~\cite{Barman:2023ymn} where a quadratic potential was assumed.
	The Hubble expansion rate is given by
	\begin{equation}
	H^2 = \frac{1}{3\, M_P^2}\,\left(\rp + \rR + \rGW\right)\,.
	\end{equation}
	Note that $(M - \Eom)/M$ and $\Eom/M$ correspond to the fractions of inflaton energy injected into SM radiation and GWs, respectively. 
	
	\subsection[Contribution to $\DNeff$]{\boldmath Contribution to $\DNeff$}
	\label{sec:DNeff}
	When the production of gravitons terminates after reheating, the GW energy density redshifts as $a^{-4}$, mimicking that of SM radiation. As a result, GW itself acts as an additional source of radiation with the potential to alter the prediction of BBN. Thus, an excess of the GW energy density around $T \sim \text{MeV}$ can be restricted by considering the (present and future) limits on $\DNeff$ from CMB and BBN. From Eqs.~\eqref{eq:beq2} and~\eqref{eq:beq3}, it follows that the evolution of the ratio GW to SM radiation energy density is
	\begin{equation}
	\frac{d(\rGW/\rR)}{da} \simeq \frac{2\, n}{2 + n}\, \frac{1}{a\, H}\, \frac{\rp}{\rR} \left[\int \frac{d\Gamma^{(1)}}{d\Eom}\, \frac{\Eom}{M}\, d\Eom - \frac{\rGW}{\rR}\, \Gamma^{(0)}\right],
	\end{equation}
	which, taking into account the scaling of $\rp$ and $T$ given in Eqs.~\eqref{eq:rpsol} and~\eqref{eq:Tevol}, can be further reduced to 
	\begin{equation} \label{eq:dRhoda}
	\frac{d(\rGW/\rR)}{da} \simeq \frac{2\, n}{2 + n}\, \frac{1}{a\, H(\Trh)}\, \left(\frac{\arh}{a}\right)^{\frac{3\, n}{2 + n} - 4\, \alpha} \left[\int_0^{\frac{M(a)}{2}} \frac{d\Gamma^{(1)}}{d\Eom}\, \frac{\Eom}{M}\, d\Eom - \frac{\rGW}{\rR}\, \Gamma^{(0)}\right],
	\end{equation}
	where, at the end of reheating, the equality $\rR(\arh) = \rp(\arh)$ is met.
	Considering the coupling of the inflaton to a pair of fermions or a pair of bosons, we obtain
	\begin{equation} \label{eq:rhoGWR}
	\frac{d(\rGW/\rR)}{da} \simeq \frac{2n}{2 + n} \frac{1}{a\, H(\Trh)} \left(\frac{\arh}{a}\right)^{\frac{3n}{2 + n} - 4\alpha} \times
	\begin{dcases}
	\frac{3 \yeff^2}{1024\pi^3} \frac{M(a)^3}{M_P^2} - \frac{\rGW}{\rR} \frac{\yeff^2}{8\pi} M(a) &\text{fermions,}\\
	\frac{1}{768 \pi^3} \frac{\mueff^2 M(a)}{M_P^2} - \frac{\rGW}{\rR} \frac{1}{8\pi} \frac{\mueff^2}{M(a)} &\text{bosons.}
	\end{dcases}    
	\end{equation}
	The previous equations can be analytically solved taking into account the scale dependence of the inflaton mass and assuming that at the beginning of reheating both the SM and GW energy densities were vanishingly small. Imposing $H(\Trh) = \Gamma^{(0)}(\Trh)$ and $\yeff^2 = 8\pi\, H(\Trh)/\Mrh$ for the fermionic and $\mueff^2 = 8\pi\, \Mrh\, H(\Trh)$ for the bosonic reheating scenarios, respectively, it follows from Eq.~\eqref{eq:rhoGWR} that
	\begin{equation} \label{eq:ror}
	\frac{\rGW}{\rR}(a) \simeq \frac{\mathcal{C}}{192 \pi^2} \frac{n}{n-3} \left(\frac{\Mrh}{M_P}\right)^2 \left(\frac{\arh}{a}\right)^\frac{6 (n-2)}{2+n} \left[\left(\frac{a}{\amax}\right)^\frac{4(n-3)}{2+n} - 1\right]\,,
	\end{equation}
	with $\mathcal{C} = 1$ for scalars and $\mathcal{C} = 9/4$ for fermions.
	\begin{figure}[t]
		\centering
		\includegraphics[scale=0.49]{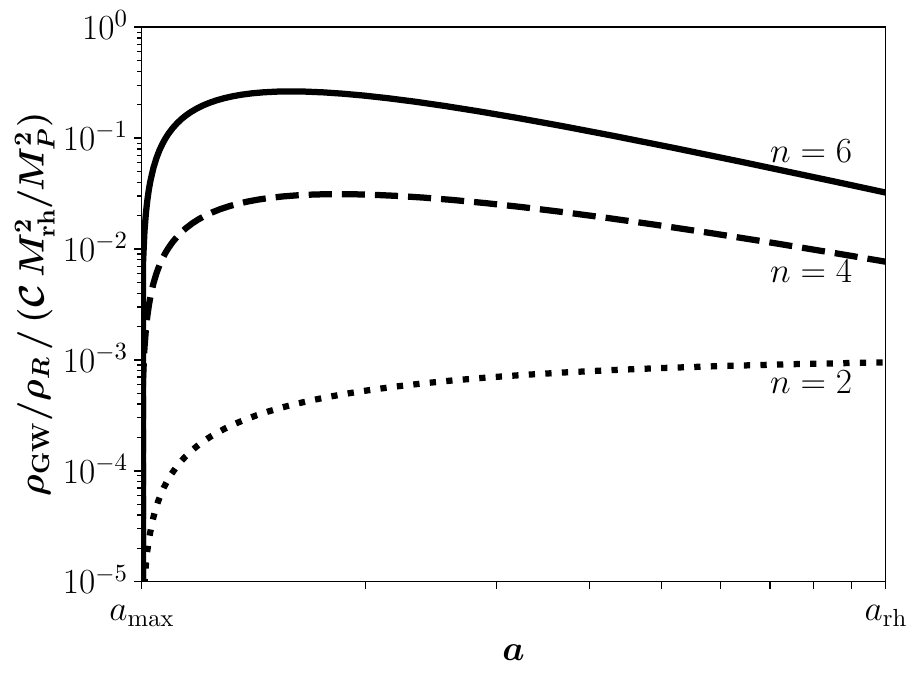}
		\includegraphics[scale=0.49]{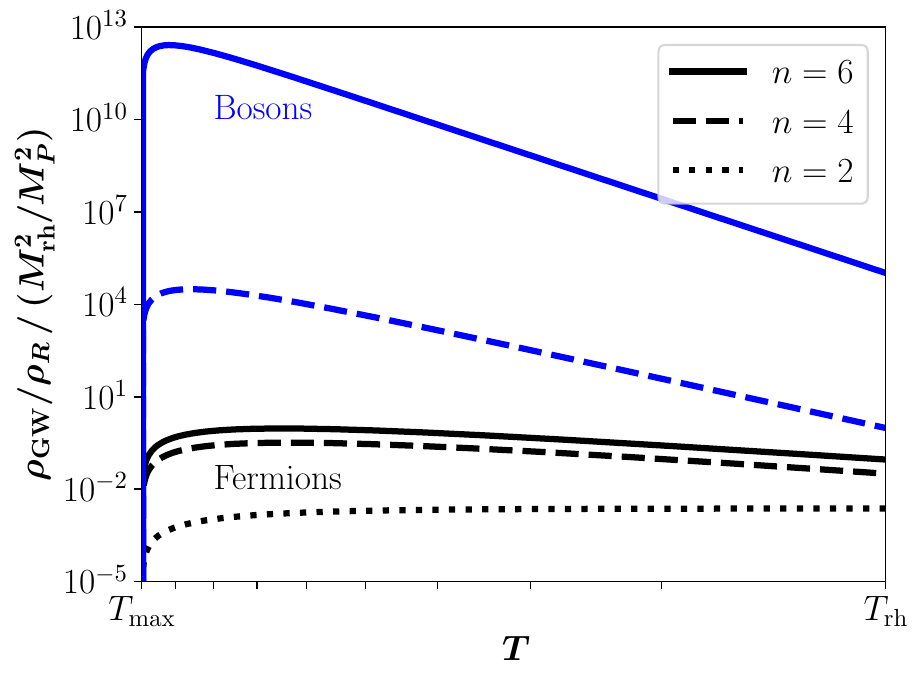}
		\caption{Evolution of the ratio $\rGW/\rR$ as a function of the scale factor $a$ (left panel) and the bath temperature $T$ (right panel) during reheating, for $\arh/\amax =10$ and $\Tmax/\Trh = 10$, respectively. On the vertical axis of the left panel $\mathcal{C} = 1\, (9/4)$ for bosonic (fermionic) reheating [cf. Eq.~\eqref{eq:ror}]. In the right panel, the blue curves correspond to bosonic reheating, while the black curves correspond to fermionic reheating. For $n = 2$ the two cases are identical.}
		\label{fig:ror}
	\end{figure} 
	In the left panel of Fig.~\ref{fig:ror} the evolution of the ratio of GW to radiation energy density $\rGW/\rR$ during reheating, as a function of the scale factor $a$ is shown, taking $\arh/\amax = 10$, for $n=2$, 4, and 6. It is interesting to observe that for $n=2$ the bulk of the GW energy density is produced at the end of reheating, near $a = \arh$. However, if $n \geq 4$, GWs are produced {\it during} reheating, close to $a \sim \amax$. The different spins in the final states give a small difference encoded in the overall factor $\mathcal{C}$. However, the impact of the spin becomes more clear when the ratio $\rGW/\rR$ is shown as a function of the bath temperature, as in the right panel of Fig.~\ref{fig:ror}. Here, the black and blue lines correspond to fermionic and bosonic reheating, respectively, for $\Tmax/\Trh = 10$. It is important to note that, in the bosonic case, the ratio $\rGW/\rR$ is larger with respect to the fermionic case because the former experiences a smaller dilution during reheating. As we will see, a larger $\rGW/\rR$ ratio shall also help to improve the GW spectrum in the bosonic scenario, compared to the fermionic reheating scenario for $n>2$.
	
	From Eq.~\eqref{eq:ror} it follows that at the end of reheating one has
	\begin{equation} \label{eq:rhoGWRTrh}
	\frac{\rGW(\Trh)}{\rR(\Trh)} \simeq \frac{n}{n-3} \left(\frac{\Mrh}{M_P}\right)^2 \times
	\begin{dcases}
	\frac{3}{256 \pi^2} \left[\left(\frac{\Tmax}{\Trh}\right)^{\frac83 \frac{n - 3}{n - 1}} - 1\right] &\text{for fermions,}\\
	\frac{1}{192 \pi^2} \left[\left(\frac{\Tmax}{\Trh}\right)^{\frac83 (n - 3)} - 1\right] &\text{for bosons,}
	\end{dcases}
	\end{equation}
	in the limit where the decay products are massless ($y \to 0$). Here we have integrated between $\amax \leq a \leq \arh$, corresponding to photon temperatures $\Tmax \geq T \geq \Trh$. During reheating, in which the SM thermal bath is produced and the universe transitions to radiation domination, the bath temperature may rise to a value $\Tmax$ that exceeds $\Trh$~\cite{Giudice:2000ex}. For $n = 2$, if $\Tmax \gg \Trh$ the square bracket tends to unity~\cite{Barman:2023ymn}, while if $\Tmax \gtrsim \Trh$, the square bracket corresponds to a reduction factor for GW production. Interestingly, it could also become an enhancement factor for $n > 3$. For example, for the case $n = 4$, we obtain
	\begin{equation}\label{eq:rho-rh}
	\frac{\rGW(\Trh)}{\rR(\Trh)} \simeq 
	\begin{dcases}
	\frac{3}{64 \pi^2} \left(\frac{\Mrh}{M_P}\right)^2 \left(\frac{\Tmax}{\Trh}\right)^\frac89 &\text{ for fermions,}\\
	\frac{1}{48 \pi^2} \left(\frac{\Mrh}{M_P}\right)^2 \left(\frac{\Tmax}{\Trh}\right)^\frac83 & \text{for bosons,}
	\end{dcases}    
	\end{equation}
	which features a power-law dependence on the ratio $\Tmax/\Trh$. The effective number of neutrinos at a temperature $T=T_{\DNeff}$ is expressed via
	\begin{equation}\label{eq:rho-dneff}
	\rho_\text{rad}(T_{\DNeff}) = \rho_\gamma + \rho_\nu + \rGW = \left[1 + \frac78 \left(\frac{T_\nu}{T_\gamma}\right)^4 N_\text{eff}\right] \rho_\gamma(T_{\DNeff})\,,
	\end{equation}
	where $\rho_\gamma$, $\rho_\nu$, and $\rGW$ correspond to the photon, SM neutrino, and GW energy densities, respectively, with $T_\nu/T_\gamma = (4/11)^{1/3}$. In particular,  Eq.~\eqref{eq:rho-dneff} is valid for $T_{\DNeff} \lesssim m_e$, that is, after electron-positron annihilation when neutrinos are decoupled from the thermal plasma. 
	
	The presence of extra relativistic degrees of freedom (in terms of GW) implies a deviation from the prediction of the SM $N_\text{eff}^\text{SM} = 3.046$~\cite{Dodelson:1992km, Hannestad:1995rs, Dolgov:1997mb, Mangano:2005cc, deSalas:2016ztq, EscuderoAbenza:2020cmq, Akita:2020szl, Froustey:2020mcq, Bennett:2020zkv}. This deviation is parameterized as
	\begin{equation}\label{eq:DNeff}
	\DNeff \equiv N_\text{eff} - N_\text{eff}^\text{SM} = \frac87 \left(\frac{11}{4}\right)^\frac43 \frac{\rGW}{\rho_\gamma}\Bigg|_{T_{\DNeff}} = \frac87 \left[\frac{11}{4} \frac{\gss(T\lesssim T_{\DNeff})}{\gss(\Trh)}\right]^\frac43 \frac{\gs(\Trh)}{2} \frac{\rGW}{\rR}\Bigg|_{\Trh}.
	\end{equation}
	Exploiting Eq.~\eqref{eq:rhoGWRTrh} we obtain
	\begin{equation}\label{eq:DNeff2}
	\DNeff \simeq \frac{11}{768 \pi^2}\, \frac{n}{n-3} \left(\frac{\Mrh}{M_P}\right)^2 \times
	\begin{dcases}
	9 \left[\left(\frac{\Tmax}{\Trh}\right)^{\frac83 \frac{n - 3}{n - 1}} - 1\right] & \text{for fermions,}\\
	4 \left[\left(\frac{\Tmax}{\Trh}\right)^{\frac83 (n - 3)} - 1\right] & \text{for bosons,}
	\end{dcases}
	\end{equation}
	where we have considered $\gss(T\simeq\text{MeV})\simeq 10.75$ and $\gss(\Trh) = \gs(\Trh) = 106.75$. For $n = 2$  we reproduce the result presented in Ref.~\cite{Barman:2023ymn} where a quadratic potential was considered. Alternatively, for $n=4$ we find
	\begin{equation}
	\DNeff \simeq
	\begin{dcases}
	5.2 \times 10^{-2}\left(\frac{\Mrh}{M_P}\right)^2 \left(\frac{\Tmax}{\Trh}\right)^{\frac89} &\text{for fermions,}\\
	2.3 \times 10^{-2}\left(\frac{\Mrh}{M_P}\right)^2 \left(\frac{\Tmax}{\Trh}\right)^{\frac83} &\text{for bosons.}
	\end{dcases}
	\end{equation}
	
	\begin{table}[t!]
		\begin{center}
			\begin{tabular}{|c||c|}
				\hline
				$\DNeff$ & Experiments  \\ 
				\hline\hline
				$0.34$ & Planck legacy data~\cite{Planck:2018vyg}\\
				\hline
				$0.14$ & BBN+CMB combined~\cite{Yeh:2022heq}\\
				\hline
				$0.06$ & CMB-S4~\cite{Abazajian:2019eic} \\
				\hline
				$0.027$ & CMB-HD~\cite{CMB-HD:2022bsz}\\
				\hline
				$0.013$ &  COrE~\cite{COrE:2011bfs}, Euclid~\cite{EUCLID:2011zbd} \\
				\hline
				$0.06$ & PICO~\cite{NASAPICO:2019thw} \\
				\hline
			\end{tabular}
		\end{center}
		\caption {Present and future constraints on $\DNeff$ from different experiments.}
		\label{tab:DNeff}
	\end{table}
	There are several present bounds and future experimental projections on $\DNeff$. Here, we briefly summarize them in Tab.~\ref{tab:DNeff}. Furthermore, as mentioned in Ref.~\cite{Ben-Dayan:2019gll}, a hypothetical cosmic-variance-limited (CVL) CMB polarization experiment could presumably be reduced to as low as $\DNeff \lesssim 3 \times 10^{-6}$. Generically, the bound of $\DNeff$ is applicable to an integrated energy density in the present epoch~\cite{Maggiore:1999vm, Caprini:2018mtu} as $\left(h^2\, \rho_\text{GW}/\rho_c\right)_0 = \int\, df/f\, \left(h^2\, \oGW(f)\right)$, where the GW spectral energy density $\oGW$ shall be derived shortly. However, with an exception for the case of a GW spectrum with a very narrow peak of width $\Delta f \ll f$, we can simply consider $\oGW^{(0)}(f)\, h^2 \leq 5.6 \times 10^{-6}\, \DNeff$, over a wide range of frequencies.
	
	\subsection{Spectrum of stochastic gravitational waves} 
	\label{sec:GWspectrum}
	Contrary to the previous case, for the GW spectrum one needs to keep track of the differential GW energy density $d\rGW/d\Eom$. Is it therefore convenient to write Eq.~\eqref{eq:beq3} in its differential form
	\begin{equation}
	\frac{d}{dt}\frac{d\rGW}{d\Eom} + 4\, H\, \frac{d\rGW}{d\Eom} = + \frac{2\, n}{2 + n}\, \frac{d\Gamma^{(1)}}{d\Eom}\, \frac{\Eom}{M}\, \rp\,.
	\end{equation}
	The evolution of the differential ratio $d(\rGW/\rR)/d\Eom$ is given by
	\begin{align} \label{eq:dGWdR}
	\frac{d}{da} \frac{d(\rGW/\rR)}{d\Eom} &\simeq \frac{2n}{2 + n} \frac{1}{a\, H} \frac{\rp}{\rR} \left[\frac{d\Gamma^{(1)}}{d\Eom} \frac{\Eom}{M} - \frac{d(\rGW/\rR)}{d\Eom} \Gamma^{(0)}\right] \nonumber\\
	&\simeq \frac{2n}{2 + n} \frac{1}{a\, H(\Trh)} \left(\frac{\arh}{a}\right)^{\frac{3n}{2 + n} - 4 \alpha} \left[\frac{d\Gamma^{(1)}}{d\Eom} \frac{\Eom}{M} - \frac{d(\rGW/\rR)}{d\Eom} \Gamma^{(0)}\right].
	\end{align}
	It is important to emphasize that in the previous expressions, $\Eom$ corresponds to the graviton energy at the moment of emission. However, to compute the GW spectrum, one needs to take into account the redshift and sum over different (redshifted) energies. Therefore, in Eq.~\eqref{eq:dGWdR} the change of variable $\Eom(\Eom',\, a) = \Eom'\, \frac{\arh}{a}$, with $\Eom'$ being the energy at $a = \arh$, is required, yielding to
	\begin{equation} \label{eq:dGWdR2}
	\frac{d}{da} \frac{d(\rGW/\rR)}{d\Eom'} \simeq \frac{\arh}{a} \frac{2n}{2 + n} \frac{1}{a\, H(\Trh)} \left(\frac{\arh}{a}\right)^{\frac{3n}{2 + n} - 4 \alpha} \left[\frac{d\Gamma^{(1)}}{d\Eom'} \frac{\Eom'}{M} - \frac{a}{\arh} \frac{d(\rGW/\rR)}{d\Eom'} \Gamma^{(0)}\right],
	\end{equation}
	where the overall factor $\arh/a$ comes from the Jacobian related to the change of variable.
	
	The primordial GW spectrum at present $\oGW(f)$ per logarithmic frequency $f$ is defined as
	\begin{equation} \label{eq:oGW}
	\oGW(f) = \frac{1}{\rho_c}\, \frac{d\rGW}{d\ln f} = \Omega_\gamma^{(0)}\, \frac{d(\rGW/\rR)}{d\ln f} = \Omega_\gamma^{(0)}\, \frac{\gs(\Trh)}{\gs(T_0)} \left[\frac{\gss(T_0)}{\gss(\Trh)}\right]^{4/3}\,\frac{d(\rGW(\Trh)/\rR(\Trh))}{d\ln \Eom'}\,,
	\end{equation}
	where $\rho_c$ is the critical energy density and $\Omega_\gamma^{(0)} h^2 \simeq 2.47 \times 10^{-5}$ is the observed photon abundance~\cite{Planck:2018vyg}.
	The GW frequency at present can be associated with the graviton energy $\Eom'$ at the end of reheating via 
	\begin{equation}
	f = \frac{\Eom'}{2 \pi}\, \frac{\arh}{a_0} = \frac{\Eom'}{2 \pi}\, \frac{T_0}{\Trh} \left[\frac{\gss(T_0)}{\gss(\Trh)}\right]^{1/3}\,,
	\end{equation}
	considering the redshift of the GW energy between reheating and the present epoch. The frequency is bounded from above because the graviton at production could carry at most half of the inflaton energy, namely $\Eom(a) \leq M(a)/2$~\cite{Barman:2023ymn}, which translates into
	\begin{equation}
	f \leq \frac{\Mrh}{4 \pi}\, \frac{\arh}{a_0} \left(\frac{\arh}{a}\right)^\frac{2 (n-4)}{n+2} \leq \frac{\Mrh}{4 \pi}\, \frac{T_0}{\Trh} \left[\frac{\gss(T_0)}{\gss(\Trh)}\right]^{1/3} \times
	\begin{dcases}
	1 &\text{ for } n \leq 4,\\
	\left(\frac{\Tmax}{\Trh}\right)^\frac{2 (n-4)}{\alpha (n+2)} &\text{ for } n > 4\,,
	\end{dcases}
	\end{equation}
	or, equivalently,
	\begin{equation}
	f \leq  4.1 \times 10^{12}~\text{Hz}\, \frac{\Mrh}{M_P}\, \frac{5.5 \times 10^{15}~\text{GeV}}{\Trh} \times
	\begin{dcases}
	1 &\text{ for } n\leq 4,\\
	\left(\frac{\Tmax}{\Trh}\right)^{\frac{4(n-4)}{3}} &\text{ for bosons } n > 4, \\
	\left(\frac{\Tmax}{\Trh}\right)^{\frac43 \frac{n-4}{n-1}} &\text{ for fermions } 4 < n < 7,\\
	\left(\frac{\Tmax}{\Trh}\right)^{\frac{2(n-4)}{n+2}} &\text{ for fermions } n > 7\,.
	\end{dcases}
	\end{equation}
	
	\subsubsection{Fermionic reheating}
	\begin{figure}[t]
		\centering
		\includegraphics[scale=0.48]{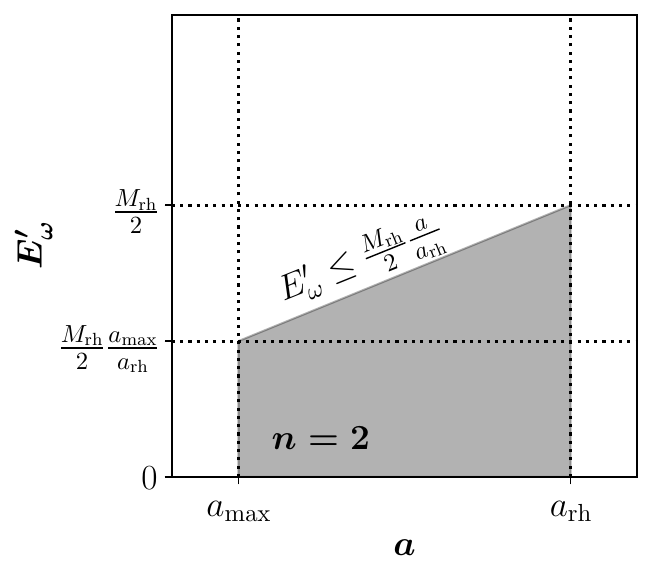}
		\includegraphics[scale=0.48]{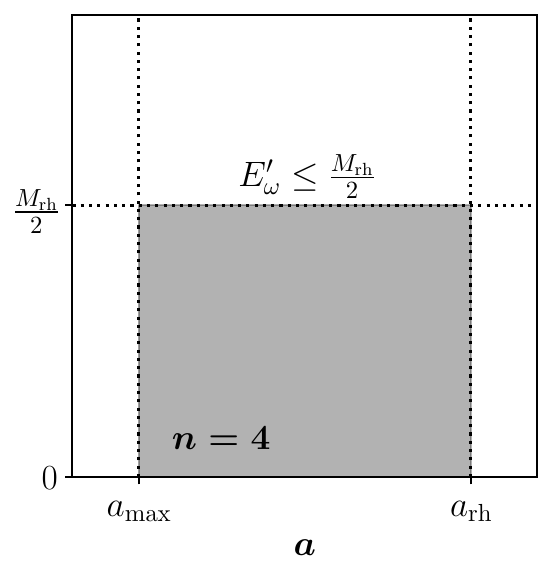}
		\includegraphics[scale=0.48]{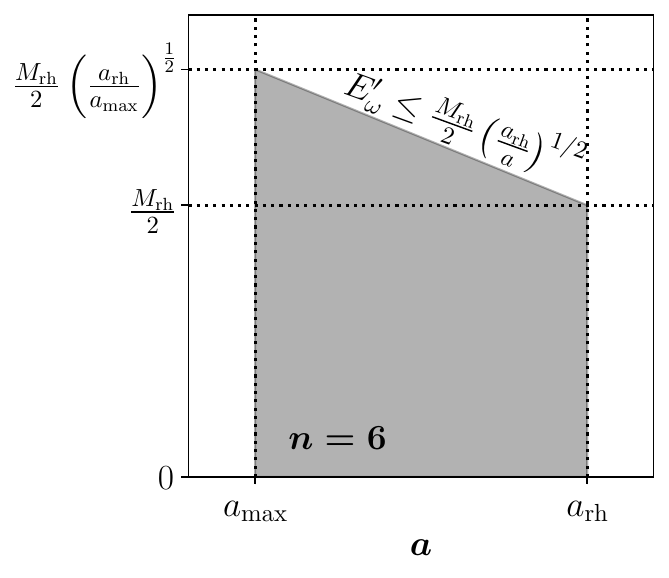}
		\caption{Schematic plot for the regime of integration in the plane $[a,\, \Eom']$, for $n=2$, 4, and 6. The vertical lines correspond to the scale factor at $a=\arh$ and $a=\amax$. }
		\label{fig:Ewp}
	\end{figure} 
	For the fermionic case, Eq.~\eqref{eq:dGWdR2} can be analytically solved. Taking into account that there is no energy stored in GWs at the end of inflation, the expression of $d(\rGW/\rR)/d\Eom'$ at the end of reheating, that is at $a = \arh$, for the case $n = 2$ is given by
	\begin{align}
	\frac{d(\rGW/\rR)}{d\Eom'} \simeq& \frac{1}{48 \pi^2} \frac{\Mrh}{M_P^2} \left[\frac{2 \Eom'}{\Mrh} \left[12 \left(1 - \left(\frac{\Tmax}{\Trh}\right)^\frac83\right) + \frac92 \frac{2 \Eom'}{\Mrh} \left(\left(\frac{\Tmax}{\Trh}\right)^\frac{16}{3} - 1\right) \right.\right.\nonumber\\
	&\quad + \left.\left. \left(\frac{2 \Eom'}{\Mrh}\right)^2 \left(1 - \left(\frac{\Tmax}{\Trh}\right)^8\right)\right] + 16 \ln \frac{\Tmax}{\Trh}\right]\,,
	\end{align}
	for $0 \leq \Eom' \leq \frac12\, \Mrh\, (\Tmax/\Trh)^{-8/3}$.
	Here, the integration has been performed between $\amax \leq a \leq \arh$. Alternatively,
	for $\frac12\, \Mrh\, (\Tmax/\Trh)^{-8/3} \leq \Eom' \leq \frac12\, \Mrh$ one has
	\begin{align}
	\frac{d(\rGW/\rR)}{d\Eom'} \simeq& \frac{1}{96 \pi^2} \frac{\Mrh}{M_P^2} \Bigg[ 2 \left( \left(\frac{2 \Eom'}{\Mrh}\right)^3 - 1\right) + 9 \left(1 - \left(\frac{2 \Eom'}{\Mrh}\right)^2\right) \nonumber\\
	&\quad + 24 \left( \frac{2 \Eom'}{\Mrh} - 1\right) - 12 \ln \frac{2 \Eom'}{\Mrh}\Bigg]\,,
	\end{align}
	after integrating between $2 \arh\, \Eom'/\Mrh \leq a \leq \arh$.
	We note that the two cases correspond to two different regimes, as shown in Fig.~\ref{fig:Ewp}. In the first case, low-energy GWs are produced during the entire reheating process, while in the second case, high-energy GWs can only be emitted during the last stage of reheating, as shown in Fig.~\ref{fig:Ewp}. This can be understood taking into account: $i)$ the maximal energy of the graviton at production is $\Eom(a) < M(a)/2$, $ii)$ for $n = 2$ the inflaton mass is constant $M(a) = \Mrh$, and $iii)$ the redshift of the GW energy.
	Alternatively, for $n = 4$ one has
	\begin{align}
	\frac{d(\rGW/\rR)}{d\Eom'} \simeq& \frac{1}{8 \pi^2} \frac{\Mrh}{M_P^2} \left[\left(\frac{\Tmax}{\Trh}\right)^\frac89 - 1\right] \nonumber\\
	&\quad \times \left[\left(1 - \left(\frac{2 \Eom'}{\Mrh}\right)^3\right) + 3 \left(\left(\frac{2 \Eom'}{\Mrh}\right)^2 - 1\right) + 4 \left(1 - \frac{2 \Eom'}{\Mrh}\right)\right]\,,
	\end{align}
	for $0 < \Eom' \leq \frac{\Mrh}{2}$. Interestingly, for $n = 4$, a single expression is required, such as the graviton energy and the inflaton mass scale in the same way: as the inverse of the scale factor $a$.
	
	For $n = 6$, on the other hand, one has
	\begin{align}
	\frac{d(\rGW/\rR)}{d\Eom'} \simeq& \frac{1}{80 \pi^2} \frac{\Mrh}{M_P^2} \left(\frac{2 \Eom'}{\Mrh}\right)^2 \Bigg[24 \ln \frac{\Tmax}{\Trh} + 15 \left(\frac{2 \Eom'}{\Mrh}\right) \left(\left(\frac{\Tmax}{\Trh}\right)^{-\frac{8}{15}} - 1\right) \nonumber\\
	&\quad + 60 \left(\frac{\Mrh}{2 \Eom'}\right) \left(1 - \left(\frac{\Tmax}{\Trh}\right)^\frac{8}{15}\right) + 15 \left(\frac{\Mrh}{2 \Eom'}\right)^2 \left(\left(\frac{\Tmax}{\Trh}\right)^\frac{16}{15} - 1\right) \Bigg]\,,
	\end{align}
	for $0 \leq \Eom' \leq \frac12\, \Mrh$, while for $\frac12\, \Mrh \leq \Eom' \leq \frac12\, \Mrh\, (\Tmax/\Trh)^{8/15}$ one has
	\begin{align}
	\frac{d(\rGW/\rR)}{d\Eom'} \simeq & \frac{3}{16 \pi^2} \frac{\Mrh}{M_P^2} \left(\frac{2 \Eom'}{\Mrh}\right)^2 \Bigg[ 3 \ln\left[\left(\frac{\Mrh}{2 \Eom'}\right) \left(\frac{\Tmax}{\Trh}\right)^\frac{8}{15}\right] + \left(\left(\frac{2 \Eom'}{\Mrh}\right) \left(\frac{\Tmax}{\Trh}\right)^{-\frac{8}{15}} - 1\right) \nonumber\\
	&\qquad + 4 \left(1 - \left(\frac{\Mrh}{2 \Eom'}\right) \left(\frac{\Tmax}{\Trh}\right)^\frac{8}{15}\right) + \left(\left(\frac{\Mrh}{2 \Eom'}\right)^2 \left(\frac{\Tmax}{\Trh}\right)^\frac{16}{15} - 1\right)\Bigg]\,.
	\end{align}
	In the latter case, the integration of Eq.~\eqref{eq:dGWdR2} has to be done in two steps: first between $\amax \leq a \leq \arh\, (\Mrh/(2 \Eom'))^2$ {\it with} the source term, then between $\arh\, (\Mrh/(2 \Eom'))^2 \leq a \leq \arh$ {\it without} the source term, as GW emission can only occur for $\Eom(a) \leq M(a)/2$. It is interesting to note that when $n = 6$ (and in general when $n \geq 6$), graviton energies higher than $\frac12 \Mrh$, up to $\Eom' \leq \frac12 \Mrh\, (\Tmax/\Trh)^{\frac2\alpha \frac{n-4}{n+2}}$, are allowed. This can be understood by noticing that the inflaton mass decreases as $M(a) \propto a^{3/2}$, i.e., faster than the energy redshift. Therefore, high-energy GWs with energies $\Eom' > \frac12 \Mrh$ can be produced at the beginning of reheating when the inflaton was very massive (heavier than its effective mass at the end of reheating).
	GWs with energies $\Eom' < \frac12 \Mrh$ can be produced during the entire reheating process.
	
	\subsubsection{Bosonic reheating}
	For the scenario where the inflaton decays into scalar final states, we follow the same procedure presented in the previous section.
	For the case $n = 2$, $d(\rGW/\rR)/d\Eom'$ at the end of reheating is given by
	\begin{align}
	\frac{d(\rGW/\rR)}{d\Eom'} \simeq \frac{1}{32 \pi^2} \frac{\Mrh}{M_P^2} \Bigg[&\left(\frac{2 \Eom'}{\Mrh}\right)^2 \left(\left(\frac{\Tmax}{\Trh}\right)^\frac{16}{3} - 1\right) - 4 \left(\frac{2 \Eom'}{\Mrh}\right) \left(\left(\frac{\Tmax}{\Trh}\right)^\frac83 - 1\right)\nonumber\\
	&\quad + \frac{16}{3} \ln \frac{\Tmax}{\Trh}\Bigg]\,,
	\end{align}
	for $0 \leq \Eom' \leq \frac12\, \Mrh\, (\Tmax/\Trh)^{-8/3}$, whereas for $\frac12\, \Mrh\, (\Tmax/\Trh)^{-8/3} \leq \Eom' \leq \frac12\, \Mrh$ one has
	\begin{equation}
	\frac{d(\rGW/\rR)}{d\Eom'} \simeq \frac{1}{32 \pi^2} \frac{\Mrh}{M_P^2} \Bigg[ \left(1 - \left(\frac{2 \Eom'}{\Mrh}\right)^2\right) + 4 \left(\left(\frac{2 \Eom'}{\Mrh}\right) - 1\right) - 2 \ln \frac{2 \Eom'}{\Mrh}\Bigg]\,.
	\end{equation}
	
	Again the $n = 4$ scenario comes with a single expression as before, that reads
	\begin{equation}
	\frac{d(\rGW/\rR)}{d\Eom'} \simeq \frac{1}{8 \pi^2} \frac{\Mrh}{M_P^2} \left(1 - \frac{2 \Eom'}{\Mrh}\right)^2 \left(\left(\frac{\Tmax}{\Trh}\right)^\frac83 - 1\right)\,,
	\end{equation}
	for $0 < \Eom' \leq \frac12 \Mrh$. Finally, for $n = 6$ one has
	\begin{align}
	\frac{d(\rGW/\rR)}{d\Eom'} \simeq& \frac{1}{32 \pi^2} \frac{\Mrh}{M_P^2} \Bigg[16 \left(\frac{2 \Eom'}{\Mrh}\right)^2 \ln \frac{\Tmax}{\Trh} \nonumber\\
	&\qquad + 3 \left(\left(\frac{\Tmax}{\Trh}\right)^\frac{16}{3} - 1\right) + 12 \left(\frac{2 \Eom'}{\Mrh}\right) \left(1 - \left(\frac{\Tmax}{\Trh}\right)^\frac83\right)\Bigg]\,,
	\end{align}
	for $0 \leq \Eom' \leq \frac12\, \Mrh$, whereas for $\frac12\, \Mrh \leq \Eom' \leq \frac12\, \Mrh\, (\Tmax/\Trh)^{8/3}$ one obtains
	\begin{align}
	\frac{d(\rGW/\rR)}{d\Eom'} \simeq& \frac{3}{32 \pi^2} \frac{\Mrh}{M_P^2} \left(\frac{2 \Eom'}{\Mrh}\right)^2 \Bigg[2 \ln\left[\left(\frac{\Mrh}{2 \Eom'}\right) \left(\frac{\Tmax}{\Trh}\right)^\frac83\right] \nonumber\\
	&\quad + 4 \left(1 - \left(\frac{\Mrh}{2 \Eom'}\right) \left(\frac{\Tmax}{\Trh}\right)^\frac83\right) + \left(\left(\frac{\Mrh}{2 \Eom'}\right)^2 \left(\frac{\Tmax}{\Trh}\right)^\frac{16}{3} - 1\right)\Bigg]\,.
	\end{align}
	
	Before closing, we note that, as a sanity check, the results of Eq.~\eqref{eq:rhoGWRTrh} for both fermions and scalars can be recovered by integrating $d(\rGW/\rR)/d\Eom'$ over the corresponding range of energy.
	
	\section{Numerical Results and Discussions}
	\label{sec: results}
	\begin{figure}[t!]
		\centering
		\includegraphics[scale=0.45]{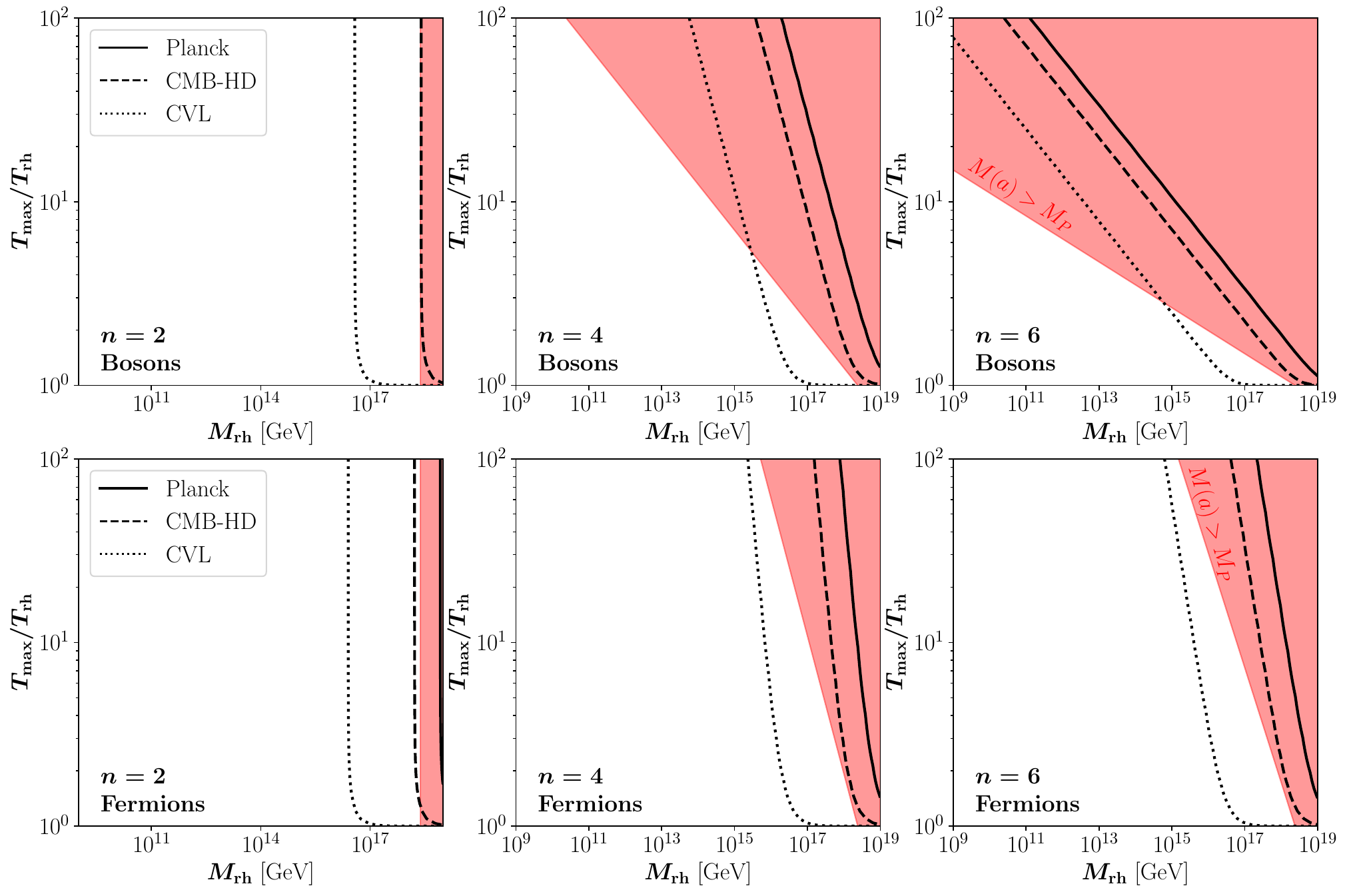}
		\caption{Contours showing the contribution of GW energy density to $\DNeff$, for $\DNeff = 3 \times 10^{-6}$ (CVL, dotted), 0.013 (CMB-HD, dashed) and 0.34 (Planck, solid), assuming $y \to 0$. Red bands are excluded since $M(a) > M_P$.
		}
		\label{fig:DNeff}
	\end{figure} 
	We show the contribution of the GW energy density to $\DNeff$, cf. Subsection~\ref{sec:DNeff}, in Fig.~\ref{fig:DNeff} for different choices of $n$, considering both bosonic (top panel) and fermionic (bottom panel) decay of the inflaton. There, we also depict the current constraint from Planck (solid lines) as well as sensitivity projections from CMB-HD (dashed lines) and the hypothetical CVL experiment (dotted lines). For each curve, the region on its right-hand side is ruled out (in the case of Planck) or could be tested in the future. However, as we see, except for CVL, the corner of the parameter space that is constrained by $\DNeff$, is already ruled out from the super-Planckian inflaton mass $M(a) > M_P$, shown by the red-shaded region. Therefore, $\DNeff$ does not impose any significant constraint on the parameter space.
	A small fraction of the parameter space is within reach of the projected sensitivity of the CVL experiment.
	
	\begin{figure}[t!]
		\centering
		\includegraphics[scale=0.86]{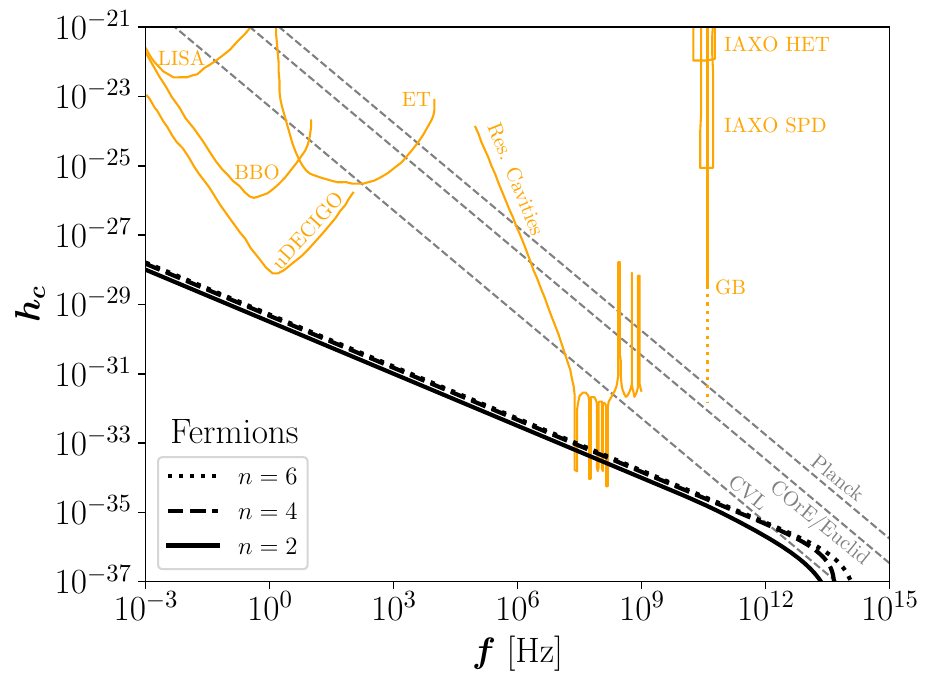}
		\includegraphics[scale=0.86]{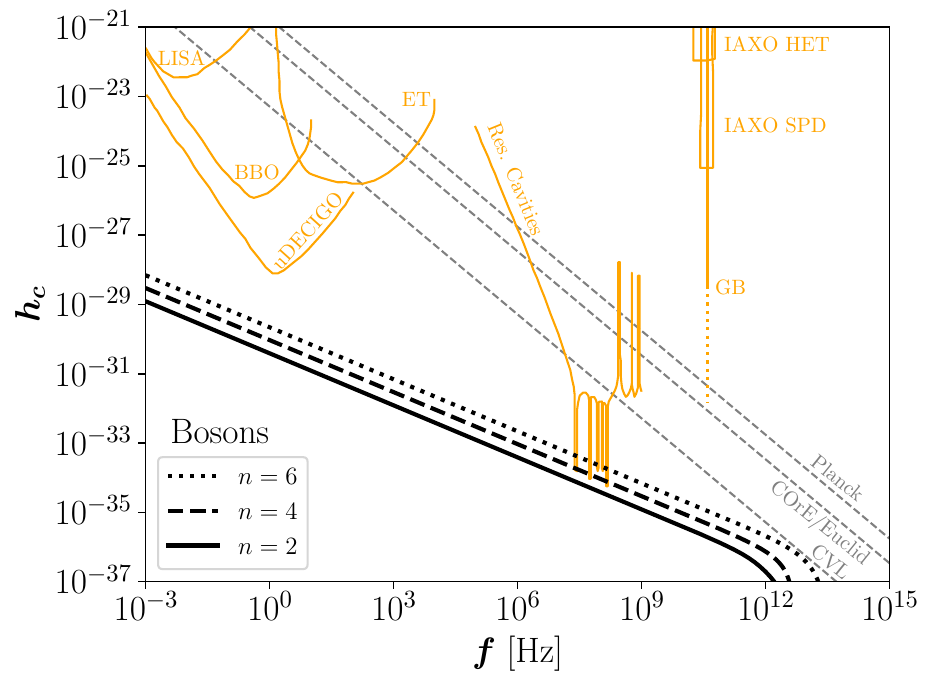}
		\caption{Dimensionless strain of GW, as a function of frequency, considering a fermionic reheating with $\Mrh = 5 \times 10^{16}$~GeV, $\Trh = 10^{13}$~GeV, and $\Tmax/\Trh = 10$ (upper panel) or a bosonic reheating with $\Mrh = 5 \times 10^{15}$~GeV, $\Trh = 10^{13}$~GeV, and $\Tmax/\Trh = 2$ (lower panel).}
		\label{fig:GWs}
	\end{figure} 
	In order to project limits from different GW experiments, it is convenient to construct the dimensionless strain $h_c$ in terms of the GW spectral energy density as~\cite{Maggiore:1999vm}
	\begin{equation}
	h_c(f) = \frac{H_0}{f}\, \sqrt{\frac{3}{2 \pi^2}\, \oGW(f)} \simeq 1.26 \times 10^{-18} \left(\frac{\rm{Hz}}{f}\right) \sqrt{h^2\, \oGW(f)}\,,  
	\end{equation}
	where $H_0 \equiv H(T_0) \simeq 1.44 \times 10^{-42}$~GeV is the present-day Hubble parameter and $h = 0.674$~\cite{Planck:2018vyg}. The upper panel of Fig.~\ref{fig:GWs} shows the dimensionless strain of GW as a function of frequency, considering fermionic reheating for $\Mrh = 5 \times 10^{16}$~GeV, $\Trh = 10^{13}$~GeV, and $\Tmax/\Trh = 10$. In this case, the increase in $n$ slightly affects the GW spectrum. If we go from $n = 2$ to $n = 4$, the spectrum is boosted by a factor $\frac34 (\Tmax/\Trh)^{8/9} / \ln(\Tmax/\Trh)$.
	Similarly, going from $n = 4$ to $n = 6$ one gets an $\mathcal{O}(1)$ factor $\sim \frac34 (\Tmax/\Trh)^{8/45}$. The lower panel of Fig.~\ref{fig:GWs} shows the dimensionless strain of GW as a function of frequency, in the case of bosonic decay, for $\Mrh = 5 \times 10^{15}$~GeV, $\Trh = 10^{13}$~GeV, and $\Tmax/\Trh = 2$. Contrary to the fermionic case, for bosons, the increase in $n$ has a significant impact on the GW spectrum. If we go from $n = 2$ to $n = 4$, the spectrum is boosted by a factor $\frac34 (\Tmax/\Trh)^{8/3} / \ln(\Tmax/\Trh)$. Similarly, going from $n = 4$ to $n = 6$ one gets a factor $\sim \frac34 (\Tmax/\Trh)^{8/3}$. Note that this choice of parameters obeys the upper bound on $\Trh$ derived in Appendix~\ref{sec:Trh} for both fermionic and bosonic cases, and the consistency relation $M_P > M(a) > T(a)$. In Fig.~\ref{fig:GWs} we also project the limits from several proposed GW detectors, for example, LISA~\cite{2017arXiv170200786A}, the Einstein Telescope (ET)~\cite{Punturo:2010zz, Hild:2010id, Sathyaprakash:2012jk, Maggiore:2019uih}, the Big Bang Observer (BBO)~\cite{Crowder:2005nr, Corbin:2005ny, Harry:2006fi}, ultimate DECIGO (uDECIGO)~\cite{Seto:2001qf, Kudoh:2005as}, GW-electromagnetic wave conversion in vacuum (solid) and in a Gaussian beam (GB) (dotted)~\cite{Li:2003tv, Ringwald:2020ist}, resonant cavities~\cite{Berlin:2021txa, Herman:2022fau}, and the International Axion Observatory (IAXO)~\cite{Armengaud:2014gea, IAXO:2019mpb} in the same plane.\footnote{The projected sensitivity curves are adapted from Refs.~\cite{Ringwald:2020ist, Ringwald:2022xif}.} Interestingly, the Bremsstrahlung-induced GWs are well within the reach of resonant-cavity detectors in the high-frequency regime. Still, at the same time, at lower frequencies, they might become sensitive to future space-based GW detectors, e.g., uDECIGO. 
	
	\section{Conclusions} \label{sec:concl}
	In this study, we investigate the production of stochastic Gravitational Waves (GWs) through graviton bremsstrahlung during the reheating period. To ensure a viable scenario, we assume that the inflaton oscillates in a generic monomial potential of the form $\phi^n$ during reheating. Unlike oscillations in a quadratic potential, this leads to a time-dependent decay width of the inflaton due to the field-dependent inflaton mass. The reheating process occurs through the inflaton's perturbative decay into either a pair of bosons or fermions, resulting in a successful reheating that satisfies the BBN measurements. As a result of the unavoidable gravitational interaction resulting from the graviton-matter coupling, the inflaton undergoes 3-body decay with the radiative emission of massless gravitons, which comprise the GW spectrum.
	
	We find that the GW spectrum is affected by the details of the reheating, or, in other words, by $i)$ the equation of state parameter of the inflaton during reheating, and $ii)$ the type of inflaton-matter coupling. Importantly, for $n > 2$, there is a significant increase in the amplitude of GW during bosonic reheating, which is still safe from the constraint of $\DNeff$ on the extra radiation energy density around BBN or CMB, in terms of GWs. However, for the fermionic reheating scenario, this boost becomes negligible for $n>4$ due to the typical functional dependence of the decay width on the inflaton mass. For a certain choice of parameters, we find, the predicted high-frequency GW signal is within the sensitivities of resonant cavity detectors for both bosonic and fermionic reheating. Our finding thus opens up the possibility that future (high-frequency) GW experiments could potentially shed light on the microscopic dynamics of reheating, for example, on the shape of the inflaton potential and types of inflaton-matter couplings. 
	
	\acknowledgments
	NB thanks Panos Oikonomou for valuable discussions. BB and YX acknowledge the illuminating discussions at the joint EISA/MITP topical workshop on ``Theoretical Particle Cosmology in the Early and Late Universe'' at Corfu, Greece. NB received funding from the Spanish FEDER / MCIU-AEI under the grant FPA2017-84543-P. YX receives support from the Cluster of Excellence ``Precision Physics, Fundamental Interactions, and Structure of Matter'' (PRISMA$^+$ EXC 2118/1) funded by the Deutsche Forschungsgemeinschaft (DFG, German Research Foundation) within the German Excellence Strategy (Project No. 39083149). OZ has received funding from the Ministerio de Ciencia, Tecnología e Innovación (MinCiencias - Colombia) through grants 82315-2021-1080 and 80740-492-2021, and has been partially supported by Sostenibilidad-UdeA and the UdeA/CODI Grant 2020-33177.
	
	\appendix
	\section{Three-body Differential Decay Rate}
	\label{sec:3body}
	The differential decay rate for the scalar final state with the emission of a graviton of energy $\Eom$ reads
	\begin{equation} \label{eq:dGdE_bos}
	\frac{d\Gamma_0^{(1)}}{d\Eom} = \frac{1}{32\,\pi^3} \left(\frac{\mueff}{M_P}\right)^2 \left[\frac{(1 - 2 x)\, (1 - 2 x + 2 y^2)}{4 x\, \xi^{-1}} + \frac{y^2\, (y^2 + 2 x - 1)}{x} \ln\left(\frac{1 + \xi}{1 - \xi}\right)\right]\,, 
	\end{equation}
	with $x \equiv \Eom/M$ and 
	\begin{equation}
	\xi \equiv \sqrt{1 - \frac{4\, y^2}{1-2x}}\,,
	\end{equation}
	with a graviton energy spanning the range
	\begin{equation}
	0 < \Eom \leq M \left(\frac12 - 2\, y^2\right)\,.
	\end{equation}
	For the fermionic final state, we obtain 
	\begin{align} \label{eq:dGdE_fer}
	\frac{d\Gamma_{1/2}^{(1)}}{d\Eom} = \frac{\yeff^2}{64\,\pi^3} \left(\frac{M}{M_P}\right)^2\Bigg[&\frac{1 - 2 x}{x\, \xi^{-1}} \left[8 x\, y^2 + 2 x\, (x - 1) - 8 y^4 - 2 y^2 + 1\right] \nonumber\\
	&+ \frac{4\,y^2 \left[(5 - 8 x)\, y^2 - (x - 1)^2 - 4 y^4\right]}{x} \ln\left(\frac{1 + \xi}{1 - \xi}\right) \Bigg]\,.
	\end{align}
	We follow the Feynman rules in Appendix~A of Ref.~\cite{Barman:2023ymn} for the computation of the decay rates.
	
	\section{Bound on the Reheating Temperature} \label{sec:Trh}
	The evolution of the radiation energy density can be extracted from Eqs.~\eqref{eq:rR_fer} and~\eqref{eq:rR_bos}, for fermionic and scalar decays, respectively.
	It follows that the thermal bath reaches a maximum temperature $T = \Tmax$, at
	$a = \amax$, when only a small fraction of the inflaton has decayed~\cite{Chung:1998rq, Giudice:2000ex}, with
	\begin{equation} \label{eq:amax}
	\amax = a_I \times
	\begin{dcases}
	\left(\frac32\, \frac{n - 1}{n + 2}\right)^\frac{n + 2}{2\, (n - 7)} & \text{ for fermions,}\\
	\left(\frac23\, (n + 2)\right)^\frac{n + 2}{2\, (2 n + 1)} & \text{ for bosons.}
	\end{dcases}
	\end{equation}
	Taking into account the upper limit on the inflationary scale $H_I^\text{CMB} \leq 2.0\times 10^{-5}~M_P$~\cite{Planck:2018jri, BICEP:2021xfz}, an upper bound on $\Trh$ can be extracted~\cite{Barman:2021ugy, Bernal:2019mhf}. For fermions, the radiation energy density at $a = \amax$ given by $\rR(\amax) = \frac{\pi^2}{30}\, \gs\, \Tmax^4$, together with Eqs.~\eqref{eq:Hubble}, \eqref{eq:rR_fer}, \eqref{eq:Tevol}, and~\eqref{eq:amax}, and the fact that $\Gp(\Trh) = H(\Trh)$ can be expressed as
	\begin{equation}
	\Trh^2 \simeq \frac{3}{\pi} \sqrt{\frac{5}{\gs}}\, \sqrt{\frac{n}{n + 2}} \left(\frac32\, \frac{n - 1}{n + 2}\right)^{\frac32 \frac{n}{7 - n}}\, M_P\, H_I \left(\frac{\Tmax}{\Trh}\right)^\frac{2\, n}{1 - n}\,,
	\end{equation}
	which corresponds to
	\begin{equation}
	\Trh \lesssim
	\begin{dcases}
	3.1 \times 10^{15}~\text{GeV} \times \left(\frac{\Tmax}{\Trh}\right)^{-2} &\text{ for } n = 2\,,\\
	3.3 \times 10^{15}~\text{GeV} \times \left(\frac{\Tmax}{\Trh}\right)^{-4/3} &\text{ for } n = 4\,,\\
	3.4 \times 10^{15}~\text{GeV} \times \left(\frac{\Tmax}{\Trh}\right)^{-6/5} &\text{ for } n = 6\,.
	\end{dcases}
	\end{equation}
	Equivalently, for scalars, one can get
	\begin{equation}
	\Trh^2 \simeq \frac{1}{\pi} \sqrt{\frac{15\, n}{\gs}} \left[\left(\frac18\right)^n \left(\frac{3}{n + 2}\right)^{1 + 5 n}\right]^\frac{1}{2 (1 + 2 n)} M_P\, H_I \left(\frac{\Tmax}{\Trh}\right)^{-2\, n}\,,
	\end{equation}
	which corresponds to
	\begin{equation}
	\Trh \lesssim
	\begin{dcases}
	3.1 \times 10^{15}~\text{GeV} \times \left(\frac{\Tmax}{\Trh}\right)^{-2} &\text{ for } n = 2\,,\\
	2.8 \times 10^{15}~\text{GeV} \times \left(\frac{\Tmax}{\Trh}\right)^{-4} &\text{ for } n = 4\,,\\
	2.5 \times 10^{15}~\text{GeV} \times \left(\frac{\Tmax}{\Trh}\right)^{-6} &\text{ for } n = 6\,.
	\end{dcases}
	\end{equation}
	
	\bibliographystyle{JHEP}
	\bibliography{biblio}
\end{document}